\documentstyle{article}
\begin{document}
\newcommand{\be}{\begin{equation}}
\newcommand{\ee}{\end{equation}}
\newcommand{\p}{\partial}
\newcommand{\bea}{\begin{eqnarray}}
\newcommand{\eea}{\end{eqnarray}}
\title{The ``Swiss cheese'' cosmological model has no extrinsic curvature
discontinuity: A comment on the paper by G.A. Baker, Jr. (astro-ph/0003152)}
\author{Charles C. Dyer \\
{\em Department of Astronomy, University of Toronto,} \\ 
{\em McLennan Labs, 60 St. George St.,} \\
{\em Toronto, Ontario, Canada M5S 3H8}
\and
Chris Oliwa \\
{\em Department of Mathematics, University of Toronto,} \\
{\em Sidney Smith Hall, 100 St. George St.,} \\ 
{\em Toronto, Ontario, Canada M5S 3G3}}
\maketitle
\begin{abstract}
Contrary to a claim, the Schwarzschild solution insertion in an expanding
universe model, the so called ``Swiss cheese'' model, does not possess an
extrinsic curvature discontinuity. We show that both the intrinsic metric and
the extrinsic curvature are continuous, and point out the error that led to
the claim.
\end{abstract}
\section{Introduction}
The ``Swiss cheese'' cosmological model is a general relativistic description
of space-time. The name refers to the fact that in this model static spherical
voids are created within a larger, time-dependent space-time. A void is 
constructed by removing the background material inside a spherical boundary
and replacing the mass by a concentration of that mass at the centre of the sphere.

Mathematically, the model is realized by the matching of a 
Friedmann-Lema\^{\i}tre-Robertson-Walker (FLRW) metric as the exterior solution,
to an exterior Schwarzschild metric as the interior solution, across a spherical
boundary. The spherical boundary stays at a fixed coordinate radius in the FLRW
frame, but changes with time in the Schwarzschild frame.

The smooth matching of two space-times across a three-surface of discontinuity
$\Sigma$ is guaranteed if the Darmois junction conditions are satisfied:
the first fundamental forms (intrinsic metrics) and the second fundamental forms 
(extrinsic curvatures) calculated in terms of the coordinates on $\Sigma$, are 
identical on both sides of the hypersurface \cite{Darmois}. The Darmois junction 
conditions allow us to use different coordinate systems on both sides of the 
hypersurface. 

The continuity of the first and second fundamental forms on a matching 
hypersurface $\Sigma$ implies the continuity of the fluid pressure on
$\Sigma$ (see e.g. \cite{Stephani}). In the case of the ``Swiss cheese'' model,
it implies a dust filled (i.e. zero pressure) FLRW space-time.

Recently, Baker \cite{Baker} claimed that the extrinsic curvatures for the 
Schwarzschild and the FLRW metrics used in the ``Swiss cheese'' model cannot be 
matched at a spherical boundary. In the following section we show that
this claim is erroneous. We prove this by explicitely constructing a smooth 
matching between the Schwarzschild and FLRW space-times across a spherical 
hypersurface. In particular, we show that both the intrinsic metrics and the 
extrinsic curvatures (henceforth often refered to as the first and second 
fundamental forms respectively) are continuous on the hypersurface. We also 
verify that the pressure is continuous as required. We conclude by indicating
the error in ref. \cite{Baker} that led to the claim. 
\section{The Matching}
The general FLRW metric can be written in spherical coordinates as
\be
ds^2=dt^2-K^2(t)\left[r^2\left(d\theta^2+\sin^2\theta d\phi^2\right)
                       +\frac{dr^2}{1-kr^2}\right]\, , 
                                 \label{eq:FLRW}
\ee
where $K(t)$ is the scale factor and $k=0, \pm 1$ the curvature constant of
space. We will show that the metric (\ref{eq:FLRW}) can be joined smoothly
on a spherical hypersurface $\Sigma$ to the Schwarzschild metric
\be
ds^2=\left(1-\frac{2M}{\rho}\right)dT^2
             -\rho^2\left(d\theta^2+\sin^2\theta d\phi^2\right)
             -\left(1-\frac{2M}{\rho}\right)^{-1}d\rho^2\, . 
                                 \label{eq:Schwarz}
\ee

The first fundamental form is the metric which $\Sigma$ inherits from the 
space-time in which it is imbedded, and may be written as 
\be
\Upsilon_{\alpha\beta}=g_{ij}\frac{\p x^i}{\p u^\alpha}
\frac{\p x^j}{\p u^\beta}\, ,
\ee
where $u^\alpha=(u^1\equiv u,\, u^2\equiv v,\, u^3\equiv w)$ is the coordinate system
on the hypersurface. Greek indices run over $1,\ldots ,3,$ while Latin indices over
$1,\ldots ,4.$

The second fundamental form \cite{Eisenhart} is defined by
\be
\Omega_{\alpha\beta}=(\Gamma^p{}_{ij}n_p-n_{i,j})
                      \frac{\p x^i}{\p u^\alpha}\frac{\p x^j}{\p u^\beta}\, ,
                         \label{eq:SFF}
\ee
where $n_a$ is a unit normal to $\Sigma,$ and $\Gamma^p{}_{ij}$ are the Christoffel
symbols. If $\Sigma$ is given by the function $f[x^a(u^\alpha)]=0$, then $n_i$ can be
calculated from
\be
n_i=-\frac{f_{,i}}{|g^{ab}f_{,a}f_{,b}|^{1/2}}\, , \label{eq:unit norm eq}
\ee
where $,i$ denotes $\frac{\p}{\p x^i}$\,. 
To avoid confusion we will denote indexed quantities associated with the FLRW and
Schwarzschild  metrics by the letters $F$ and $S$ respectively.

We consider a spherical hypersurface $\Sigma$ given by the function 
$f_F(x_F^i)=r-r_0=0,$ where $r_0$ is a constant, and parametrized by 
$x_F^1=t=u,\, x_F^2=\theta=v,\, x_F^3=\phi=w,$ and $x_F^4=r=r_0,$ in the FLRW frame. 
In the Schwarzschild frame we choose the parametrization 
$x_S^1=T=T(u),\, x_S^2=\theta=v,\, x_S^3=\phi=w,$ and $x_S^4=\rho=\rho(u).$
The condition $\Upsilon_{F\alpha\beta}=\Upsilon_{S\alpha\beta}$ then implies   
\be
1=\left(1-\frac{2M}{\rho}\right)\left(\frac{dT}{du}\right)^2
    -\left(1-\frac{2M}{\rho}\right)^{-1}\left(\frac{d\rho}{du}\right)^2\, ,
                          \label{eq:First FF1}
\ee
\be
K^2r_0^2=\rho^2\, .       \label{eq:First FF2}
\ee

We now turn to the two second fundamental forms. The (outward pointing) 
unit normal in the FLRW frame can be calculated from eq. (\ref{eq:unit norm eq}) 
and $f_F(x_F^i)=r-r_0=0.$ The result is $n_{Fi}=\delta_i^4n_{F4},$ where
$n_{F4}=-|g_{F44}|^{1/2}.$ The unit normal is spacelike, i.e. $n_F^i n_{Fi}=-1$. 
The unit normal in the Schwarzschild frame cannot be obtained directly from 
eq. (\ref{eq:unit norm eq}) since we do not know the form of $f_S.$ 
However, $n_{Si}$ must satisfy the two conditions
\be
n_S^i n_{Si}\equiv n _F^i n_{Fi}=-1\;\; \mbox{  and  } \;\;
n_{Si}\frac{\partial x_S^i}{\partial u^\alpha}=0\, ,
                               \label{eq:two cond}
\ee
where the second condition results from the partial differentiation
of $f_S[x_S^i(u^\alpha)]=0$ with respect to $u^\alpha$. 
From (\ref{eq:two cond}) one obtains
\be
\left.
 \begin{array}{c}
 n_{S2}=n_{S3}=0, \\
 n_{S1}\frac{dT}{du}+n_{S4}\frac{d\rho}{du}=0, \\
 \left(1-\frac{2M}{\rho}\right)^{-1}{n_{S1}}^2
      -\left(1-\frac{2M}{\rho}\right){n_{S4}}^2=-1.
 \end{array}
\right\}                \label{eq:normal S cond}
\ee
With the help of eq. (\ref{eq:First FF1}), equations (\ref{eq:normal S cond})
enable us to derive $n_{Si}$ as a function of $u^\alpha:$
\be
n_{Si}=\left(\epsilon\frac{d\rho}{du}, \; 0, \; 0, \;
       -\epsilon\frac{dT}{du}\right)\, , \;\;\; 
        \epsilon=\pm 1\, .   \label{eq:normal K}
\ee

Because of the simple form of $n_{Fi},$ eq. (\ref{eq:SFF}) for the second 
fundamental form can be much simplified in the FLRW frame. In this case one 
obtains from (\ref{eq:SFF})
\bea
\Omega_{F\alpha\beta}
 & = & \Gamma^4_{Fij}n_{F4}
        \frac{\p x^i_F}{\p u^\alpha}\frac{\p x^j_F}{\p u^\beta}
        -n_{F4,j}\frac{\p x^4_F}{\p u^\alpha}\frac{\p x^j_F}{\p u^\beta}
                                      \nonumber \\
 & = & \Gamma^4_{F\mu\nu}n_{F4}
        \frac{\p x^\mu_F}{\p u^\alpha}\frac{\p x^\nu_F}{\p u^\beta}\, ,\ \ \ \
        \mbox{since}\ \frac{\p x^4_F}{\p u^\alpha}=\frac{\p r_0}{\p u^\alpha}=0\, ,
                                      \nonumber \\
 & = & \Gamma^4_{F\mu\nu}n_{F4}\delta^\mu_\alpha\delta^\nu_\beta \nonumber \\
 & = & n_{F4}\Gamma^4_{F\alpha\beta} \nonumber \\
 & = & -\frac{1}{2}|g_{F44}|^{1/2}g_F^{4i}
       (g_{F\alpha i,\beta}+g_{F\beta i,\alpha}-g_{F\alpha\beta,i})\nonumber \\
 & = & \frac{1}{2}|g_{F44}|^{1/2}g_F^{44}g_{F\alpha\beta,4}\, ,
\eea
so that
\be
\Omega_{F\alpha\beta}=-\frac{1}{2}|g_{F44}|^{-{1/2}}
                         g_{F\alpha\beta,4}\, .\label{eq:Simple SFF}
\ee
Equation (\ref{eq:Simple SFF}) with $F'$s dropped, i.e.
\be
\Omega_{\alpha\beta}=-\frac{1}{2}|g_{44}|^{-{1/2}}
                         g_{\alpha\beta,4}\, ,\label{eq:GenSimple SFF}
\ee
is valid for any coordinate hypersurface $x^4=\mbox{constant},$ in an
orthogonal coordinate system and parametrized by $x^\alpha=u^\alpha.$ No 
similar simplification of the second fundamental form of $\Sigma$ is possible
in the Schwarzschild frame. Moreover, for us to calculate the second term in
eq. (\ref{eq:SFF}) we need $n_{Si}$ as a function of $x_S^i.$ However, if we 
once more differentiate the second condition in eq. (\ref{eq:two cond}) with 
respect to $u^\alpha$ it follows that
$$n_{Si,j}\frac{\p x_S^i}{\p u^\alpha}\frac{\p x_S^j}{\p u^\beta}=
          -n_{Si}\frac{\p ^2 x_S^i}{\p u^\alpha\p u^\beta}\, ,$$
giving
\be
\Omega_{S\alpha\beta}=\Gamma^p_{Sij}n_{Sp}
                     \frac{\p x_S^i}{\p u^\alpha}\frac{\p x_S^j}{\p u^\beta}
                     +n_{Si}\frac{\p ^2 x_S^i}{\p u^\alpha\p u^\beta}\, .
                        \label{eq:SFF new}
\ee 
Using equations (\ref{eq:Simple SFF}) and (\ref{eq:SFF new}), we find that
$\Omega_{S\alpha\beta}=0=\Omega_{F\alpha\beta}\, ,\,\forall\,\alpha\neq\beta.$
The remaining, diagonal components of $\Omega_{\alpha\beta}$ and the 
continuity condition $\Omega_{S\alpha\beta}=\Omega_{F\alpha\beta}$ on $\Sigma,$
result in the following three differential equations 
\bea
\Omega_{S11} &\equiv & \Gamma^4_{S11}n_{S4}\left(\frac{dT}{du}\right)^2
                        +\Gamma^4_{S44}n_{S4}\left(\frac{d\rho}{du}\right)^2
                        +2\Gamma^1_{S14}n_{S1}\frac{dT}{du}\frac{d\rho}{du}
                                                 \nonumber \\
             &       &  \ +n_{S1}\frac{d^2T}{du^2}+n_{S4}\frac{d^2\rho}{du^2}
                        =0\equiv\Omega_{F11}\, , \label{eq:SFF1}\\
\Omega_{S22} &\equiv & \Gamma^4_{S22}n_{S4}=\alpha|K|r_0
                        \equiv\Omega_{F22}\, ,   \label{eq:SFF2}\\
\Omega_{S33} &\equiv & \Gamma^4_{S33}n_{S4}=\alpha|K|r_0\sin^2\theta
                        \equiv\Omega_{F33}\, ,   \label{eq:SFF3}
\eea
where the Christoffel symbols are given by 
\bea
\Gamma^1_{S14} & = & -\Gamma^4_{S44}=\frac{M}{\rho(\rho-2M)}\, ,\nonumber \\
\Gamma^4_{S11} & = & \frac{(\rho-2M)M}{\rho^3}\, ,\nonumber \\
\Gamma^4_{S22} & = & -(\rho-2M)\, ,\nonumber \\
\Gamma^4_{S33} & = & \sin^2\theta\Gamma^4_{S22}\, , \nonumber
\eea
and $\alpha\equiv (1-kr_0^2)^{1/2}.$

Equations (\ref{eq:SFF2}) and (\ref{eq:SFF3}) being equivalent, we can use 
either of them and eq. (\ref{eq:First FF2}) to obtain
\be
\frac{dT}{du}=\frac{\epsilon\alpha\rho}{\rho-2M}\, .
                \label{eq:time der}
\ee
It then follows from eq. (\ref{eq:First FF1}) that
\be
\left(\frac{d\rho}{du}\right)^2=\alpha^2-\left(\frac{\rho-2M}{\rho}\right)\, .
                           \label{eq:rho der}
\ee
Differentiating eqs. (\ref{eq:time der}) and (\ref{eq:rho der}) w.r.t. $u$, 
one obtains
\be
\frac{d^2T}{du^2}=-\epsilon\alpha\frac{2M}{(\rho-2M)^2}\frac{d\rho}{du}\, .
\ee
and
\be
\frac{d^2\rho}{du^2}=-\frac{M}{\rho^2}\, . \label{eq:rho 2der}
\ee
With equations (\ref{eq:time der})-(\ref{eq:rho 2der}), eq. (\ref{eq:SFF1}) is 
now identically satisfied. Thus, both the first and second fundamental forms 
are continuous on $\Sigma.$

It remains to verify that the pressure is also continuous across the spherical
boundary. Since in the Schwarzschild space-time the pressure is zero, it must
vanish in the FLRW space-time. The FLRW space-time is a perfect fluid 
space-time, and as such its Einstein field equations are given by
\be
   G^{ij}\equiv R^{ij}-\frac{1}{2}Rg^{ij}
   =8\pi\left[(\mu+p)u^iu^j-pg^{ij}\right]\, ,\label{eq:Einst ten}
\ee
where $G^{ij}$ is the Einstein tensor, $R^{ij}$ the Ricci tensor, $R$ the Ricci
curvature scalar, $\mu$ the matter-energy density, $p$ the pressure, and $u^i$
the unit four-velocity. Specifically, eq. (\ref{eq:Einst ten}) implies
\be
 {G^1}_1=\frac{3(\dot{K}^2+k)}{K^2}=8\pi\mu\, \label{eq:Einstein1} 
\ee 
and 
\be
 {G^2}_2={G^3}_3={G^4}_4=\frac{2\ddot{K}}{K}
                             +\frac{(\dot{K}^2+k)}{K^2}=-8\pi p\, ,
                                              \label{eq:Einstein2} 
\ee 
where $\dot{K}\equiv \frac{dK}{du}\equiv\frac{dK}{dt},$ and 
$\ddot{K}\equiv \frac{d^2K}{du^2}\equiv\frac{d^2K}{dt^2}.$

To obtain $\dot{K}$ and $\ddot{K},$ we differentiate eq. (\ref{eq:First FF2}) 
w.r.t. $u.$ The results are
\be
\dot{K}=\eta\frac{1}{r_0}\frac{d\rho}{du}\, ,\label{eq:K dot}
\ee
and
\be
\ddot{K}=\eta\frac{1}{r_0}\frac{d^2\rho}{du^2}\, ,\label{eq:K ddot}
\ee
where $\eta\equiv\frac{K}{|K|}.$
Substituting eqs (\ref{eq:rho der}) and (\ref{eq:rho 2der}) into 
(\ref{eq:K dot}) and (\ref{eq:K ddot}), and using
$K=\eta\frac{\rho}{r_0}$ (from eq. (\ref{eq:First FF2})), it follows from
(\ref{eq:Einstein2}) that $p=0.$ Finally, we note that, from 
eq. (\ref{eq:Einstein1}), $8\pi\mu=\frac{6M}{r_0^3|R|^3},$ a positive 
quantity, which shows that the matching is physically admissible.
\section{Discussion and Conclusions}
We have shown, in a mathematically rigorous way, that an FLRW space-time can 
be joined smoothly to a Schwarzschild space-time across a spherical boundary, 
a construction used in the ``Swiss cheese'' cosmological model of the Universe. 
The boundary has a fixed coordinate radius in the FLRW frame, but changes with
time in the Schwarzschild frame. In particular, we have shown that the 
intrinsic metric and extrinsic curvature are both continuous on the boundary
as is required for a smooth, permanent matching. These matching requirements
also imply the continuity of the pressure on the boundary, which has been 
verified here.

Our results differ from those of ref. \cite{Baker}. Specifically, we have not
found a discontinuity in the extrinsic curvature on the spherical boundary as
claimed in \cite{Baker}. The claim in \cite{Baker} can be explained out as
follows. The author, without justification, uses equation 
(\ref{eq:GenSimple SFF}) (his eq. (2.10) with $N_\alpha=0$) for calculating 
the components of $\Omega_{\alpha\beta}$ in both the FLRW and Schwarzschild 
frames. This results in the mismatch of $\Omega_{S\alpha\beta}$ and 
$\Omega_{F\alpha\beta}$ on the spherical hypersurface, leading to the claim 
that ``the 'Swiss cheese' model is at best only an approximation, with a 
singular interface''. 

The use of eq. (\ref{eq:GenSimple SFF}) in the Schwarzschild frame is 
incorrect. Assuming, {\em a priori,} (as the author of ref. \cite{Baker} and 
we have done) that the boundary surface changes with time in the Schwarzschild 
frame, it follows that the boundary is not a coordinate boundary in that frame, 
and, therefore, eq. (\ref{eq:SFF}) for the extrinsic curvature, cannot be 
reduced to eq. (\ref{eq:GenSimple SFF}). Equation (\ref{eq:GenSimple SFF}) can 
be used for calculating $\Omega_{S\alpha\beta}$ only under the assumption that 
the matching hypersurface in the Schwarzschild frame is the coordinate 
hypersurface $\rho=\mbox{constant.}$ But then eq. (\ref{eq:First FF2}) cannot 
be satisfied unless $R(t)=\mbox{constant;}$ obviously not an interesting case, 
since then the FLRW space-time reduces to the Minkowski space-time. Thus the 
use of eq. (\ref{eq:GenSimple SFF}) in the Schwarzschild frame is an error.
\end{document}